\documentclass[twocolumn,prl,showpacs]{revtex4}
\usepackage{graphicx,epsfig,amssymb}

\begin{document}
%\draft
%\preprint{}

\title{Calculation of dephasing times in closed quantum dots}
 
\author{Eli Eisenberg, Karsten Held, and Boris L. Altshuler}
\affiliation{Physics Department, Princeton University, Princeton, NJ 08544\\
 NEC Research Institute, 4 Independence Way, Princeton, NJ 08540 }

\begin{abstract}

Dephasing of one-particle states in closed quantum dots is analyzed 
within the framework of random matrix theory and Master equation. 
Combination of this analysis with recent experiments on 
the magnetoconductance allows for the first time to evaluate the 
dephasing times of closed quantum dots. These dephasing times turn out
to depend on the mean level spacing and to be significantly enhanced as 
compared with the case of  open dots. Moreover, the experimental data 
available are consistent with the prediction that the dephasing of
one-particle states in finite closed systems disappears at low enough 
energies and temperatures. 
\end{abstract}

\date{Version 5, \today}

\maketitle

Quantum coherence of electrons in closed quantum dots has
attracted much interest in recent years
\cite{Sivan94,Altshuler97,Sivan94b,
Folk00,Beenakker00}.
%, general interest to understand dephasing
%and particular interest based on the consideration to use quantum dots
% as qubits. 
Electron-electron interactions are believed to be one of
the main sources of dephasing in disordered systems
at low temperatues. Compared to low-dimensional metals and 
semiconductors
\cite{Altshuler85},
substantial modifications
of this dephasing mechanism are caused by 
the confinement of the quantum dot which leads to 
discrete energy levels. In particular, the
 dephasing rate was predicted \cite{Altshuler97}
to disappear  at low excitation energies,
$\epsilon<\Delta\sqrt{g/\ln g}$ where
$\Delta$ is the mean level spacing and $g$ is the dimensionless 
conductance of the dot. 

Whereas there are a number of ways to measure the dephasing times in open
quantum dots \cite{Huibers98,Clarke95}, 
the situation is much more complicated in closed dots.
Only a few experiments have attempted to study dephasing
in closed quantum dots. Most of these have focused on the relaxation
of highly excited states, verifying the continuous to discrete spectrum 
transition at $\epsilon \propto g \Delta$ \cite{Sivan94b}. 
Some signatures of 
dephasing in thermalized states have been studied by
 Patel {\it et al.} \cite{Patel98}, who
analyzed  the statistical distribution of the 
conductance maxima $G^{\rm max}$
(the height of the Coulomb blockade peaks). They found that the ratio of 
standard deviation  to mean peak height $\sigma(G^{\rm max})/\langle
G^{\rm max}\rangle$
is smaller than what random matrix theory (RMT) predicts
\cite{Alhassid98}, and 
attributed this reduction to dephasing effects. 
More recently, Folk {\it et al.} \cite{Folk00} 
suggested to use the dependence of the  conductance
upon applying a magnetic field $B$, 
\begin{equation}
\alpha =(\langle G^{\rm max} \rangle_{B\ne 0}-\langle G^{\rm max}
\rangle_{B=0})/\langle G^{\rm max} \rangle_{B\ne 0},
\label{alpha}
\end{equation}
as a probe of dephasing times. This  is the closed dot analog of
the weak localization magnetoconductance
which was analyzed earlier for open dots \cite{Huibers98}.
RMT predicts $\alpha =1/4$ \cite{Jalabert92,Alhassid01}, while
Folk {\it et al.}  measured considerably lower values of $\alpha$,
down to $\alpha \approx 0$ for the largest quantum dot with the 
maximal ratio $k_B T/\Delta$ ($T$ is the temperature, $k_B$  the 
Boltzman constant). 
%It was then suggested that 
%other dephasing mechanisms are present in addtion to electron-electron
interactions.
%Beenakker {\it et al.}  
Beenakker {\it et al.}
\cite{Beenakker00} theoretically analyzed the situation in which  the 
phase-breaking inelastic relaxation rate $\Gamma_{\rm in}$ \cite{note1} far
exceeds the mean tunneling rate (inverse dwell time in the dot) 
$\overline \Gamma$. It turns out that, in this limit, $\alpha$ is 
reduced much stronger than what found experimentally. Thus,  
%Beenakker {\it et al.} \cite{Beenakker00} 
they concluded that  in the experiment \cite{Folk00}
 $\Gamma_{\rm in} < \overline{\Gamma}$.
However, as noted in Refs. \cite{Folk00} and \cite{Beenakker00}, the lack
of a quantitative theory of the crossover regime
$\Gamma_{\rm in}\sim\overline\Gamma$, prevents a full analysis of the 
experimental results.

In this Letter, we study theoretically 
the effect of arbitrary inelastic scattering
on the conductance of a closed quantum dot. We develop 
an analytical approach that allows to evaluate 
$\alpha$ [Eq.~(\ref{alpha})] and compare the results to  the 
numerical solution. 
 The approximate results are found to reasonably 
describe the behavior 
in the experimentally relevant temperature regime.
Our calculations allow for the first time to extract dephasing
times of low lying (thermaly excited) states
in closed quantum dots. We observe a clear enhancement 
of the dephasing 
times relative to earlier results for open quantum dots \cite{Huibers98}.
Moreover, contrary to  the analysis of open quantum dots \cite{Huibers98}
which showed a dependence on temperature alone, we find a dependence 
on {\em both}  $T$ and $\Delta$. From our analysis it follows that
 the measurements of
Folk {\it et al.} \cite{Folk00} 
are not inconsistent with vanishing dephasing rate for
low excitation energies \cite{Altshuler97}.
A more detailed presentation of the calculation
will be given in \cite{Eisenberg01}.

In the experimentally relevant regime 
$\hbar \Gamma_{\rm in},\hbar \overline{\Gamma} <k_B T,\Delta$, 
each state of the quantum dot is determined by a
tuple $\{n_i\}$ of occupation numbers for the single particle 
eigenstates with energies $E_i$ and spins $S_i$.
The electrons can tunnel between the dot 
and the two leads. The left (L) and right (R) leads differ due to
the applied voltage $V$. The elctrons in the leads are thermalized
and distributed according to the Fermi function $f_{FD}(E)= 
[1+e^{E/(k_B T)}]^{-1}$.
%$\lambda \in\{L,R\}$
%which differ by an applied voltage $V$, where
%$f_j^{\lambda}= 
%[1+e^{ (E_j+ [\delta_{\lambda, L}-1/2] e V-{\mu})/(k_B T)}]^{-1}$
%is the Fermi function for the  energy an electron has which tunnels
% from state $j$ to lead $\lambda$. We denote by ${\mu}$ the 
%effective chemical potential, which includes the charging energy.
The probability $P_{\cal N}(\{n_i\})$ to find a given 
set of occupation numbers
$\{n_i\}$ with a total of  ${\cal N}$ electrons
(restricted to ${\cal N}\in\{N,N+1\}$ due to the Coulomb blockade) 
obeys the following Master equation 
\cite{Beenakker91}:

\begin{widetext}

\begin{eqnarray}
 \! \! \!{\frac{{\rm d }P_N(\{n_i\})}{{\rm d}t}} \! &\!\!=\!\!&\! \sum_{j
\lambda}\!  
\delta_{n_j 0} \Gamma_j^{\lambda}[ (1\!-\!\!f_j^{\lambda}) 
P_{N+1}(\{n_i\}_{+ \!j}) \! -\!f_j^{\lambda} P_N(\{n_i\})]
+\! \sum_{jk} \!  \delta_{n_j 0} \delta_{n_k 1} \big[\Gamma_{\rm in}^{j 
k}  
P_{N}(\{n_i\}_{+ \!j \!- \!k}) -\Gamma_{\rm in}^{k  j}P_N(\{n_i\})\big]
\label{Master1} \\
 \! \! \!{\frac{{\rm d} P_{N+1}(\{n_i\})}{{\rm d}t}} \!&\!\!=\!\!&\!
 \sum_{j \lambda} \!
\delta_{n_j 1} \Gamma_j^{\lambda} [f_j^{\lambda} P_N(\{n_i\}_{-\! j})
\!-\! (1\!-\!\!f_j^{\lambda}) P_{N+1}(\{n_i\})]
+\!\! \sum_{jk} \!  \delta_{n_j 0} \delta_{n_k 1} \big[\Gamma_{\rm
in}^{j k}  
P_{N\!+\!1}(\{n_i\}_{+ \!j \!- \!k})\! -\!\Gamma_{\rm in}^{kj}
P_{N\!+\!1}(\{n_i\})\big] 
\nonumber
\end{eqnarray}

\end{widetext}
Here,  $\{n_i\}_{+j}$ ($\{n_i\}_{-j}$) are the tuples obtained
from   $\{n_i\}$ by adding (removing) one electron in the
one-particle eigenstate $j$, and $f^\lambda_j=f_{FD}(E_j+
(\delta_{\lambda L}-1/2)eV-\mu)$, where $\lambda\in\{L,R\}$ and
$\mu$ is the effective chemical potential, including the charging energy.

The first terms in (\ref{Master1}) describe 
the tunneling of electrons between the dot and the 
leads.
The additional terms $\Gamma_{\rm in}^{k  j}$ in (\ref{Master1}) 
describe inelastic scattering processes
 between the dot's one-particle  eigenstates $j$ and $k$.
We assume that these are
caused by thermal bosonic fluctuations at temperature $T$
 and neglect any back-coupling of the
scattering to this bose bath. 
Under the assumptions
that the coupling 
strength is independent of the specific levels 
involved and there is no spin-scattering, 
one arrives at ($\omega_{jk}=E_j-E_k$)
\begin{eqnarray}
\Gamma_{\rm in}^{j k} =  \Gamma_{\rm in}^0 \; \frac{{\rm
sgn}(\omega_{jk})\; 
D(|\omega_{jk}|)}{\exp[\omega_{jk}/(k_B T)]-1}\;\delta_{S_j S_k},
\label{Gin}
\end{eqnarray}
where $D(E)$ is the bosonic density of states.
As will be shown below, the suppression of $\alpha$ is quite 
robust to the specific model of interaction, and depends mainly on
the {\em total inelastic scattering rate} $\Gamma_{\rm in}$.
We consider $\Gamma_{\rm in}$
% or  $\Gamma_{\rm in}^0$ 
as a phenomenological parameter, to be determined experimentally.
Since the experimental quantum dots \cite{Folk00} 
contain a large number of electrons,
$N\gg 1$, they  can be described by RMT \cite{Aleiner01}.
 In particular, the tunneling rates are Porter Thomas distributed
$P_{\beta}(\Gamma) \propto {\Gamma^{\beta/2-1}} \exp[-\beta \Gamma/(2
\overline{\Gamma})]$,
with $\beta=1,2$ for the Gaussian orthogonal (no magnetic field) 
and unitary ensemble (with a magnetic field). The difference between
$P_1(r)$ and $P_2(r)$ leads to the 
afore mentioned of $\alpha=1/4$ 
in the absence of inelastic scattering \cite{Jalabert92,note}.

The inelastic scattering model (\ref{Gin}) is 
exponentially cut off to states outside an energy
window of  ${\cal O}(k_BT)$ and, thus, $\Gamma_{\rm in}$ vanishes
at low temperatures.
At   $k_BT \gg \Delta$ on the other hand, there are many states 
$M\propto T/\Delta$  connected by the inelastic scattering. Therefore, 
for $T\to\infty$, the total inelastic scattering 
rate  $\Gamma_{\rm in}/\overline{\Gamma} \rightarrow \infty$
and the  result of \cite{Beenakker00} is approached.
In the leading order in ${\Delta}/{k_B T}$ the solution of the 
Master equation (\ref{Master1}) is
the equilibrium distribution and 
\begin{equation}
G=\frac{e^2}{2k_BT}  \frac{k_B T}{\Delta}
\frac{\overline{\Gamma}^L\overline{\Gamma}^R}
{\overline{\Gamma}^L+\overline{\Gamma}^R},
\label{G00}
\end{equation}
i.e., $\alpha=0$ (here, $\overline{\Gamma}^{\lambda}$ 
is the mean tunneling rate to
lead $\lambda$, $\overline{\Gamma}=\overline\Gamma^L+\overline\Gamma^R$).
Two
corrections arise
in the next order in  ${\Delta}/{k_B T}$:
(i) the effect of a finite total inelastic scattering rate
and (ii)  $\overline{\Gamma}$ is replaced by
an average $M$ levels 
around the Fermi energy, i.e.,
$\overline{\Gamma}^{\lambda}\rightarrow \langle\! \langle{\Gamma}^{\lambda}_j
\rangle \!\rangle=\sum_j f_j (1-f_j) {\Gamma}_j^\lambda$
\cite{Beenakker00,Beenakker91} (Here and in the following $f_j$ is the Fermi
function at {\it both} leads for $V=0$). This introduces corrections
${\cal O}(1/M)$ due to correlations between the nominator and denominator
of Eq. (\ref{G00}).
We calculate the former and take into account the latter
 by solving the Master equation using
perturbation theory in $\overline{\Gamma}/\Gamma_{\rm in}$ where
$\Gamma_{\rm in}^j=\sum_{k\neq j} \Gamma_{\rm in}^{j k}(1-f_k)$ 
%at $E_j={\mu}$.
As a result \cite{Eisenberg01}
\begin{eqnarray}
%P(\{n_i\})&=&P(\{n_i\})_{\rm eq} \big(1+\frac{eV}{kT} \big[\frac{\langle \langle \overline{\Gamma}^L-\overline{\Gamma}^R \rangle\rangle}{\langle \langle\overline{\Gamma}^L+\overline{\Gamma}^R \rangle \rangle}$
%\nonumber\\&&
%-\sum_j \frac{1-f_j}{2}\frac{\Gamma_j^L-\Gamma_j^R}{\Gamma_in} \big),
%\\&&
\!\!\!G &\!\!=\!\!& \frac{e^2}{k_B T} P^{\rm eq}(N) \left(\frac{\langle
\langle \Gamma^L 
\rangle \rangle \langle \langle \Gamma^R \rangle \rangle}{\langle\!
\langle 
\Gamma^L+\Gamma^R \rangle \rangle}- 
\right. \nonumber \\&& \left.
\frac{\langle\! \langle 
{\Gamma^L}^2 \rangle\!\rangle\langle\! \langle 
\Gamma^R \rangle\!\rangle^2\!+\!\langle\! \langle 
{\Gamma^R}^2 \rangle\!\rangle\langle\! \langle 
\Gamma^L \rangle\!\rangle^2\!-\!2\langle\! \langle 
\Gamma^L \rangle\!\rangle^2\langle\! \langle 
\Gamma^R \rangle\!\rangle^2}{\Gamma^*_{\rm in}\langle\! \langle 
\Gamma^L+ \Gamma^R\rangle\!\rangle^2 }\!\right)\label{GhighT}\\
\!\!\!\alpha&\!\!=\!\!&\frac{1}{12}\frac{\Delta}{k_B
T}+\frac{\overline{\Gamma}}
{2 \Gamma_{\rm in}}.
\end{eqnarray}
where $P^{\rm eq}(N)$ is the equilibrium
probability to have $N$ electrons in the dot and  
$\Gamma^*_{\rm in}=\Gamma_{\rm in}^j/(1-f_j)$. We neglect
the weak $j$-dependence of $\Gamma^*_{\rm in}$. 
% The total inelastic scattering rate does depend on energy. 
The total inelastic scattering rate $\Gamma_{\rm in}$ is defined as
value of $\Gamma_{\rm in}^j$ at Fermi energy, i.e., $\Gamma_{\rm in}=
\Gamma_{\rm in}^j$ for $E_j=0$.

In the following, we employ an approximation inspired by this 
high-T expansion.
To first order in voltage, one can write  
$P_{\cal N}(\{n_i\})=P^{\rm eq}_{\cal N}(\{n_i\}) 
\big(1+\frac{eV}{k_B T}\sum_j \delta_{n_j 1} \Psi(j) \big)$,
where $P^{\rm eq}_{\cal N}(\{n_i\})$ is the equilibrium value of 
$P_{\cal N}(\{n_i\})$.
With a large number of final
states to scatter to, one can replace the sum over many individual 
terms $P(\{n_i\}_{+k-j})$
in the Master equation (\ref{Master1}) by an averaged quantity
%\begin{widetext}\begin{equation}
%\sum_k\Gamma_{\rm in}^{q  p}  P_{N\!+\!1}(\{n_i,k|j\})- \Gamma_{\rm in}^{j k}
%  P_{N\!+\!1}(\{n_i\})}
%=\frac{eV}{k_B T} P^{\rm eq}_{N\!+\!1}(\{n_i\}) \sum_k \Gamma_{\rm in}^{j k}
%[\Psi(s)-\Psi(p)]
% \frac{eV}{k_B T} P^{\rm eq}_{N\!+\!1}(\{n_i\})
% \Gamma_{\rm in}
% \bar{\Psi}.
%\end{equation}\end{widetext}
\begin{eqnarray}
\lefteqn{\sum_k\Gamma_{\rm in}^{k j}  
P_{N\!+\!1}(\{n_i\}_{+k-j})- \Gamma_{\rm in}^{j k} 
P_{N\!+\!1}(\{n_i\})}&&\nonumber\\
 && =\frac{eV}{k_B T} P^{\rm eq}_{N\!+\!1}(\{n_i\}) \sum_k \Gamma_{\rm
in}^{j k}
[\Psi(k)-\Psi(j)]
\nonumber\\ &&\rightarrow
 \frac{eV}{k_B T} P^{\rm eq}_{N\!+\!1}(\{n_i\})
 \Gamma_{\rm in}
 [\bar{\Psi}-\Psi(j)].
\label{approx}
\end{eqnarray}
Here,  $\bar{\Psi}$ should, in principle, be a weighted average
over levels  within a range of ${\cal O}(T)$ around a particular level
$j$
considered. However, only levels around the Fermi energy
are of interest for the conductance since
the contribution of every level $j$ to the conductance is multiplied 
by $f_j(1-f_j)$. For this reason, we approximately 
employ a constant $\bar{\Psi} =\sum_j f_j(1-f_j) 
{\Psi(j)}/(\sum_j f_j(1-f_j))$  in Eqs. (\ref{approx}) and (\ref{Master1})
This leads to a self-consistent solution of the Master equation with the
result
\cite{Eisenberg01}
\begin{eqnarray}
\!\!G &\!\!=\!\!&\! \frac{e^2}{k_B T} P^{\rm eq}(N) \!
\left\langle\!\!\!\left\langle {\Gamma^L_i\tau^{\rm tot}_i}
\biggl (\!\Gamma^R_i\! + \!
%\right.\right.\nonumber\\&& \left.\left.\times
\frac{\Gamma^*_{\rm in}\big\langle\!\!\big\langle  
{\Gamma^R_j}{\tau^{\rm tot}_j}
\big\rangle \!\!\big\rangle}
{\big\langle\!\!\big\langle (\Gamma^L_j\!+\!\Gamma^R_j)
{\tau^{\rm tot}_j}\big\rangle\!\!\big\rangle}\!\biggr )\!\!
 \right\rangle\!\!\!\right\rangle \label{Gapprox}
\end{eqnarray}
where $\tau^{\rm tot}_i\!\!=\!\!(\Gamma^L_i+\Gamma^R_i+\Gamma^*_{\rm
in})^{-1}$.
One would obtain the same form (\ref{Gapprox})
but with $\langle\!\langle ... \rangle\!\rangle= \sum_{j=1}^M ...$
considering $M$ degenerate levels 
filled with $N\in\{0,1\}$ electrons.

The result  (\ref{Gapprox}) can be interpreted
in the following way: the first term represents 
processes in which the electron was not scattered at all. These happen
with 
probability $({\Gamma^L_i+\Gamma^R_i})\tau^{\rm tot}_i$
and the resulting conductance peak heights are proportional to
$\frac{\Gamma^L_i\Gamma^R_i}
{\Gamma^L_i+\Gamma^R_i}$; yielding  $ \Gamma^L_i
\Gamma^R_i\tau^{\rm tot}_i$ altogether. The second term represents
contributions from electrons that were 
inelastically scattered after tunneling from one lead, 
and their contribution to conductance is
$
\langle\!\langle {\Gamma^R_j}\tau^{\rm tot}_j \rangle\!\rangle
/\langle\!\langle
 {(\Gamma^L_j+\Gamma^R_j)}{\tau^{\rm tot}_j}\rangle\!\rangle$.

\begin{figure}[tb]
\includegraphics[width=3.25in]{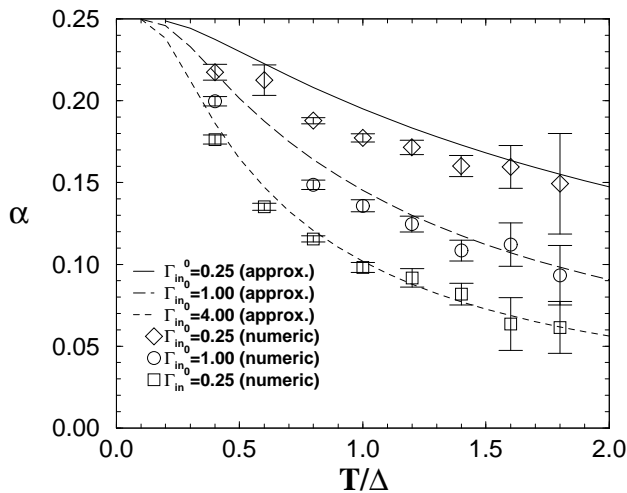}
\caption{Comparison of the numerical solution of the full Master equation
with the high temperature approximation. The lattter is seen to work
well for $k_B T > \Delta$.} 
\label{numerics}
\end{figure}
\begin{figure}[tb]
\includegraphics[width=3.25in]{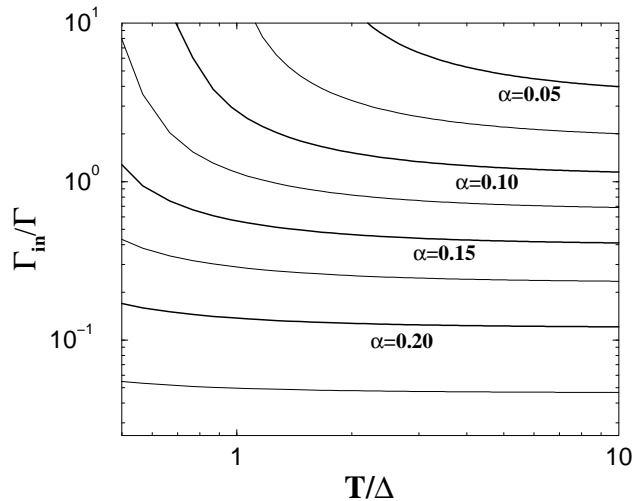}
\caption{A contour plot of $\alpha$ as a function of $T/\Delta$ and
$\Gamma_{\rm in}/\overline\Gamma$, based on the high temperature
approximation. 
The values the bold contours are specified.
Given $T,\Delta$ and $\alpha$ from future experiments, 
one can extract $\Gamma_{\rm in}/\overline\Gamma$ from this Figure.}
\label{contplot}
\end{figure}

Equation (\ref{Gapprox}) is the main result of this paper. It is based on
an approximation (\ref{approx}) which can be justified in the high 
temperature limit.
The particular advantage of this approach is that it gives not only 
the correct leading high temperature behavior [Eq.~(\ref{GhighT})] but
also
reproduces correctly the limits $\Gamma_{\rm in}=0$ 
and $\Gamma_{\rm in}=\infty$ for all $T$ including $\alpha=1/4$ 
at $T=0$. Below we demonstrate that this approach works pretty well in the
intermediate regime $k_BT\sim\Delta$.

In order to calculate 
$G$ and $\alpha$ one has  to average Eq. (\ref{Gapprox}) 
w.r.t. the different ensembles. One could do so numerically, but
it is possible to get analytical results via expanding Eq.
(\ref{Gapprox})
in powers of $\Delta/k_BT$ \cite{Eisenberg01}.  
The first three terms in the $\Delta/k_BT$ expansion already give
good accuracy in the relevant regime $k_B T>\Delta$
and are employed in the following. As we are interested
in this regime we assumed a picket fence distribution with  spacing
$\Delta$ between consecutive spin-degenerate levels
($E_{2j}=E_{2j-1}=j \Delta$; $\Gamma^{\lambda}_{2j}=
\Gamma^{\lambda}_{2j-1}$;
$\overline{\Gamma}^{\lambda}_{j}=\overline{\Gamma}/2$).

We tested the range of validity of this high-temperature approximation
against
the  numerical solution \cite{numerics} 
of the Master equations (\ref{Master1}). The latter is
obtained by solving the
Master equation (\ref{Master1}) 
by sparse matrix inversion \cite{numerics}.
Figure \ref{numerics} compares values of $\alpha$, as 
calculated using the first three terms in the  $\Delta/T$ expansion, 
with the numerical values. 
The agreement is very good for sufficiently
high temperatures, and reasonable
even for low $T$. In the whole temperature regime, the
deviations
are within current experimental accuracy.

It, thus, appears that our analytical approach 
provides a reliable way to determine
$\Gamma_{\rm in}$ from the experimental measurements of $\alpha$, 
in the whole temperature regime. For future experiments we provide Fig. 
\ref{contplot}, which
presents $\alpha$ as a contour-plot in the space spanned by 
$k_BT/\Delta$ 
and $\Gamma_{\rm in}/{\overline \Gamma}$.

\begin{figure}[tb]
\includegraphics[width=3.25in]{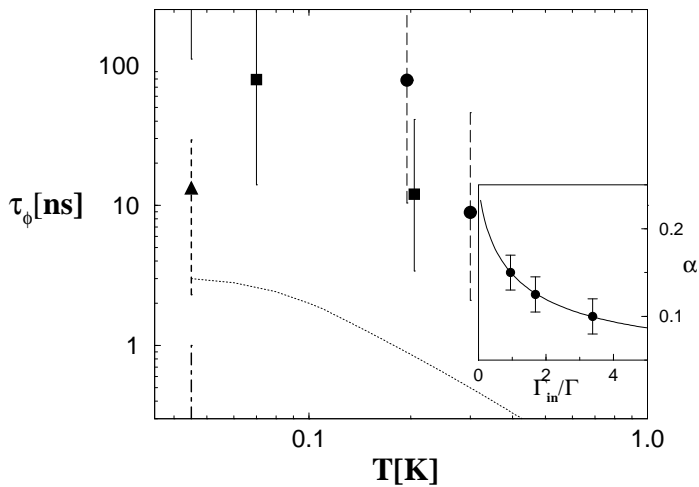}
\caption{Dephasing times, $\tau_\phi$, as extracted from the data points
in 
Ref.~\cite{Folk00} for four different dots:
$\Delta=28\mu$eV (circles, long-dashes error bars), $\Delta=10\mu$eV
(squares,
solid error bars), $\Delta=2.4\mu$eV (up-triangles, dashed error bar),
and
$\Delta=0.9\mu$eV (dot-dashed error bar); dotted line:  fit so
open dot experiments as calculated in
\cite{Huibers98}. Error bars which extend up (down)
beyond the graph should be understood as going up to infinity (down to
zero);
if  no corresponding point is visible  the experimental mean value
 itself gives $\tau_\phi=\infty$ 
(or $\tau_\phi=0$). In the inset, we fit experimental measurements for
different values of $\overline\Gamma$ \cite{Folk00} with our theory. The
single
fitting parameter is $\Gamma_{\rm in}=0.25\mu$eV, or $\tau_\phi=16$ns. }
\label{comp}
\end{figure}

A direct experimental test would be provided by
measuring values of $\alpha$ in a given dot at  fixed  $T$, 
as a function of $\overline\Gamma$ (which can be achieved by changing the
contact setting). The theoretical 
dependence of $\alpha$ on $\overline\Gamma$ involves a single fitting
parameter, i.e., the unknown total scattering rate
 $\Gamma_{\rm in}$ which is assumed to be unaffected by the contact
setting. 
A first step in this direction was done in \cite{Folk00},
and in the inset of Fig \ref{comp} we compare the 
prediction of our high temperature approximation 
with the measurements of $\alpha$ for three
different values of $\overline\Gamma$. An excellent agreement is obtained,
though more data points are required for  reliable conclusions.

We now use the above theory to extract  dephasing times
from the data points (mean values and error bars)
of Folk {\it et al.} \cite{Folk00}.
% (one standard deviation).
Figure \ref{comp} presents these estimates as symbols and error bars, 
respectively,
and compares
them  with open dot values \cite{Huibers98}. A clear
 enhancement of the dephasing times compared to 
 open dots is observed. In addition, dephasing times  strongly depend on 
$\Delta$ (as can be seen at $T=45$~mK). This is 
in contrast to open dot results 
\cite{Huibers98}.
%Two data points from Ref. \cite{Folk00} were not included, as 
%they are consistent with any value of inelastic scattering rate.
An additional suppression of $\alpha$ for $k_BT<\Delta$,
resulting from level-spacing fluctuations \cite{Eisenberg01,note} 
was not included in our analysis, 
and therefore our results {\it underestimates} 
the dephasing times for $k_BT<\Delta$.
%, i.e., the points in Fig.\ref{comp} which are  consistent even 
%with $\tau_{\phi}=\infty$ and their error bars lie even higher. 
In addition, the result
for the $\Delta =0.9 \mu$eV
quantum dot which is consistent with $\tau_{\phi}=0$
should be interepreted carefully since the result implies
$\hbar \Gamma_{\rm in} > \Delta$ and the Master equation is not
applicable anymore.
%At low temperatures, the main intrinsic source of dephasing 
%is due to electron-electron interactions \cite{xx}. 
%The dephasing rate due to electron electron interactions
%is predicted to be enhanced in a closed dot due to the confinement
%\cite{xx}, and to diverge at a critical temperature 
%$T_*\sim \Delta\sqrt{g/\ln g}$ \cite{xx}.
%In addition, several other dephasing mechanisms have been proposed
%in order to explain the saturation in the dephasing rates observed 
%in open dots. 
Based on our analysis, the recent experiment \cite{Folk00}, 
measuring dephasing in closed quantum dots is consistent
with dephasing due to electron-electron interaction alone, including
the prediction of the critical vanishing of dephasing rate. 
However, given the large error bars of the current experimental
data, one can not exclude an algebraic behavior or even a saturation 
of the dephasing rates for $T\rightarrow 0$. Nevertheless, the behavior 
is clearly different from that of open quantum dots \cite{Huibers98} 
and is $\Delta$-dependent.

In conclusion, we provide a theoretical approach to extract the 
inelastic scattering rate in closed dots based on measurements of the
weak-localization correction $\alpha$. Analyzing a recent experiment
by Folk {\it et al.} \cite{Folk00},
we see a clear enhancement of the dephasing time compared with
open dots values. There is no  inconsistency 
with theoretical predictions for electron-electron
interaction, in particular, a vanishing dephasing rate  at a critical 
$\Delta-dependent$ temperature.
We note, however, that the available experimental data is limited and
has considerably statistical uncertainties.
Future experiments are necessary and we offer
Figure \ref{contplot} to extract  the  temperature and
level-spacing dependence of the inelastic scattering rate and 
to thoroughly test the prediction of a diverging dephasing time.

\begin{acknowledgments}
We are happy to acknowledge intensive and very helpful discussions
with Igor Aleiner, Joshua Folk, and Charles Marcus. This work has
been supported by ARO, DARPA, and the Alexander von Humboldt 
foundation.
\end{acknowledgments}

%\end{references}


\begin{thebibliography}{10}
\bibitem{note}
In the temperature regime $0.1 \Delta < T<\Delta$, differences in 
the GOE/GUE level spacing distributions 
lead to a significant reduction from the value $\alpha=1/4$ 
(down to $\alpha \approx 0.2$) \cite{Eisenberg01}. This effect alone
can explain the suppression of $\alpha$ in the experiment \cite{Folk00} 
for  $k_BT<\Delta$.
\bibitem{numerics} For $k_B T<1.6\Delta$, we took into account 
all configurations
$\{ n_i\}$ which involve levels in $[-4 k_B T\, .. \,4 k_B T]$ with 
$P^{\rm eq}\{ n_i\}/\min_{\{ n_i\}} P^{\rm eq}\{ n_i\}> \exp(-5\Delta/k_B
T)$;
for $k_B T>1.6\Delta$ the interval was  $[-3 k_B T\, ..\,3 k_B T]$
and the exponential cut-off $-4.5\Delta/k_B T$, see \cite{Eisenberg01} for
details.
\end{thebibliography}
\end{document}